\documentclass[11pt,epsf,letterpaper]{article}%
\usepackage{color}
\usepackage{amsmath}
\usepackage{amsfonts}
\usepackage{verbatim}
\usepackage{amssymb}
\usepackage{graphicx}
\usepackage{mathrsfs}%
\providecommand{\U}[1]{\protect\rule{.1in}{.1in}}
\begin{document}

\title{Asymptotically (anti) de Sitter Black Holes and Wormholes with a Self
Interacting Scalar Field in Four Dimensions.}
\author{Andr\'{e}s Anabal\'{o}n$^{1}$ and Adolfo Cisterna$^{2}$.\\$^{1}$\textit{Departamento de Ciencias, Facultad de Artes Liberales y}\\\textit{Facultad de Ingenier\'{\i}a y Ciencias, Universidad Adolfo
Ib\'{a}\~{n}ez, Vi\~{n}a del Mar, Chile.}\\$^{2}$\textit{Instituto de F\'{\i}sica, Pontificia Universidad de Cat\'{o}lica
de Valpara\'{\i}so,}\\\textit{Av. Universidad 330, Curauma, Valpara\'{\i}so, Chile.}}
\maketitle

\begin{abstract}
The aim of this paper is to report on the existence of a wide variety of exact
solutions, ranging from black holes to wormholes, when a conformally coupled
scalar field with a self interacting potential containing a linear, a cubic
and a quartic self interaction is taken as a source of the energy-momentum
tensor, in the Einstein theory with a cosmological constant. Among all the
solutions there are two particularly interesting. On the one hand, the
spherically symmetric black holes when the cosmological constant is positive;
they are shown to be everywhere regular, namely there is no singularity
neither inside nor outside the event horizon. On the other hand, there are
spherically symmetric and topological wormholes that connect two
asymptotically (anti) de Sitter regions with a different value for the
cosmological constant. The regular black holes and the wormholes are supported
by everywhere regular scalar field configurations.

\end{abstract}

\section{Introduction and discussion.}

The interplay of scalar fields and the gravitational interaction has a long
and rich history. Although it dates backs to the Brans-Dicke article
\cite{Brans:1961sx}, it is fair to say that there was a large increment in the
interest with Wheeler's no hair conjecture \cite{RW}. It states that is not
possible to endow nor to deform a four dimensional asymptotically flat black
hole, regular on and outside the horizon with a scalar field that is regular
in the above mentioned region; the domain of outer communications plus its
boundary to be precise. This conjecture has been proved in a number of cases
(for references and a list of cases where the no hair conjecture is a theorem
see \cite{AyonBeato:2002cm}) and for real scalar fields in four dimensions
there is no much doubt that it is true; a nice account of this history can be
found in \cite{Bekenstein:1996pn}.

As time went by theoretical and observational arguments moved the community to
take the inclusion of a cosmological constant seriously. Allowing the
spacetime to be asymptotically of constant curvature changed the picture and a
number of black holes were found \cite{Duff:1999rk}-\cite{Kolyvaris:2011fk}.
The no hair conjecture, in these cases, was therefore recast as a no primary
hair\footnote{The scalar hair is called primary when there is an extra
integration constant associated to it, otherwise it is called secondary.}
conjecture for black holes of spherical topology and a scalar field potential
derivable from a superpotential \cite{Hertog:2006rr}. It is worth to mention
that nowadays there is a renewed interest in the no hair conjecture; high
precision astronomical observation of the supermassive black holes has been
argued to be a way to experimentally test it, see for instance
\cite{Sadeghian:2011ub}.

Scalar fields are also present in the standard cosmological model \cite{PDG},
in any compactification and therefore in most of the extended supergravity
theories. As has been previously mentioned the existence of black holes with
scalar fields in anti de Sitter spacetime is, now, widely known. The fact that
these hairs are of secondary kind, namely that there is no integration
constant associated to the scalar field, implies that it is not possible to
continuously connect the hairy configuration with mass $M$ and a configuration
with the same mass and no scalar field. This gives rise to a phase transition
at constant free energy in the gravity theory \cite{Martinez:2004nb}. This
kind of second order phase transition has become well known as the gravity
dual of a superconductor, for references and a review see
\cite{Hartnoll:2009sz}.

Here a family of wormholes that are asymptotically of constant curvature is
constructed in this paper. Anti de Sitter wormholes have gained attention
within the holographic context and they have been thoroughly analyzed when the
energy momentum tensor vanishes \cite{Witten:1999xp}. Since the existence of
these kind of objects was disproved when the boundary is globally within the
conformal equivalence class of $R\times S^{N}$, later it was studied whether
wormholes with an hyperbolic horizon would be of interest for the AdS/CFT
conjecture \cite{Maldacena:2004rf}. Holography was analyzed in a five
dimensional wormhole spacetime in \cite{Arias:2010xg}. In a more general
context these spacetimes have been analyzed within higher dimensional
Chern-Simmons theory \cite{Dotti:2006cp}, Horava gravity
\cite{BottaCantcheff:2009mp}, conformal gravity \cite{Oliva:2009zz}, Lovelock
gravity \cite{Canfora:2008ka} and non miminimally coupled electrodynamics
\cite{Balakin:2007am, Balakin:2010ar}, as well as evolving Lorentzian
wormholes \cite{Cataldo:2012pw} just to mention a few . To add an example in a more simple theory a
family of asymptotically anti de Sitter hairy wormhole solutions are
constructed in this paper within the four dimensional Einstein theory when the
boundary is $R\times S^{2}$ or $R\times H^{2}$.

The wormholes also exist when the cosmological constant is positive. Each of
the asymptotic regions have a different value of the cosmological constant and
also a different value for the scalar field, corresponding to a different
extreme of the potential. In the de Sitter case it is also possible to
eliminate the cosmological horizon and the solution becomes an inhomogeneous,
anisotropic, cosmology. It starts from a completely homogenous state, with
positive constant curvature $\frac{\lambda}{\xi^{2}}$, evolves to an
inhomogeneous state and ends again in a completely homogenous de Sitter space
but with a different value for the cosmological constant $\lambda$, $\xi$
being an arbitrary parameter of the scalar field potential. It follows that
the solution can interpolate between an arbitrarily large cosmological
constant in the past and a very small one in the future (or vice versa).

It is also interesting to remark that the black holes have no inner
singularity, an issue rather studied when the spacetime is asymptotically
flat, and where no example for scalar fields with positive kinetic energy is
known, although there are explicit examples when a non linear electrodynamic
theory is included \cite{AyonBeato:2004ih} (for references and a deeper
discussion on regular black holes see the previous reference), for some
numerical results in the same directions see \cite{Nucamendi:1995ex}. In this
paper the inner singularity of the black holes is replaced by another
asymptotic region, with a different value for the cosmological and the
gravitational constant.

The study of the backreaction of scalar fields on the spacetime also brings in
an interesting perspective regarding the dark matter issue. Originally the
dark matter problem was related to the impossibility of fitting the orbital
velocities of galaxies in clusters with the Newtonian potential, $\Phi
=-\frac{GM}{r}$, $M$ being the visible matter within the cluster. The simplest
explanation would therefore be that $M$ has another source for which nowadays
many candidates exist \cite{Mukhanov:2005sc}. When dark matter was proposed
the cosmological constant was considered a mathematical curiosity and
therefore, at the level of the Einstein equations, the requirement of
asymptotic flatness made the assumption on the form of the gravitational
potential rather reasonable. However, the inclusion of the cosmological
constant allows a larger variety of potentials that can be dominant between
the relevant scale of the Newtonian potential and the relevant scale of the
cosmological constant. Namely, all the functions that grow slower than $r^{2}$
in the asymptotic region and grow slower than $r^{-1}$ close to the surface of
the star or black hole. Actually, an upshot of the theoretical studies
associated to the detailed analysis of the asymptotic behavior of scalar
fields in asymptotically anti de Sitter spacetimes, has been the form of the
explicit contribution to the total energy of the spacetime coming from the
slow fall off of the scalar field and its backreaction on the metric
\cite{Henneaux:2004zi}.

In our Universe, better modeled by an asymptotically de Sitter spacetime, the
metric can very well fit the observed galaxy rotation curves due to the
modification of the gravitational potential. Indeed, since for spherically
symmetric solutions in the Schwarzschild gauge, the gravitational potential is
related to the metric as $g_{00}=-1-2\Phi$. The presence of a scalar field, or
any other particle with slow fall off, would give rise to a new kind of
potential and therefore the amount of dark matter present in a given model
should be reconsidered. Moreover, if dark matter can be modeled as a field and
it fills the spacetime from the center of the galaxy all the way to the dark
matter halo a simple candidate to analyze this situation is a scalar field. It
has already been pointed out that a gravitational potential, arising from a scalar field model \cite{Nucamendi:2000jw} can give account of the galaxy rotation curves. If the dark matter component of the galaxy comes from a new integration constant, arising from a scalar field, it would give a
very simple explanation to the fact that the amount of dark matter in every
galaxy is different\footnote{Since the mass is an integration constant, within
the general relativity realm, there is no fundamental explanation for its
value in any gravitating object. If the dark matter content of a gravitating
system would also be related to an integration constant no fundamental
explanation would be necessary to give account of the difference in the dark
matter content of every galaxy.} (ranging from 1\% to 99\%). In this paper a
three parametric gravitational potential is shown to arise when the
backreaction of a conformally coupled scalar field with a polynomial potential
is considered. Instead of having three integration constants it only has one,
the usual mass, plus two parameters of the action principle, namely the
cosmological constant and a parameter coming from the scalar potential. We
believe that a new integration constant in spherically symmetric
configurations would necessarily imply a sort of primary hair which could come
either from the matter action or having a pure gravitational origin as in
\cite{Anabalon:2011bw}.

The outline of the paper is as follows: in the first section the model is
presented, the solution is explicitly written and some important features of
it remarked. As usual, due to the non-linearity structure of the field
equations the solution has been constructed by an educated guess, therefore no
reference to its derivation is made. In the second section, the geometrical
characterization of the solutions is done. In the third section, further
remarks and comments are done. The notation follows~\cite{wald}. The
conventions of curvature tensors are such that an sphere in an orthonormal
frame has positive Riemann tensor and scalar curvature. The metric signature
is taken to be $(-,+,+,+)$, $(\partial\phi)^{2}=g^{\mu\nu}\partial_{\mu}%
\phi\partial_{\nu}\phi$, $\delta_{\lambda\rho}^{\mu\nu}=\delta_{\lambda}^{\mu
}\delta_{\rho}^{\nu}-\delta_{\rho}^{\mu}\delta_{\lambda}^{\nu}$. Greek letters
are in the coordinate tangent space and latin letters in the non-coordinate
tangent space, $8\pi G=\kappa$ and the units are such that $c=1=\hbar$.

\section{The model and the solutions.}

As is well known, the Green function of a massless scalar field have support
on the light cone on Minkowski spacetime. To extend this behavior to a curved
background, such that the Green function goes to its flat space form when the
curvature goes to zero, the massless scalar field must be conformally coupled
to the scalar curvature \cite{Sonego:1993fw} (for an interesting
generalization of the conformally coupled scalar field to higher dimensions
see \cite{Oliva:2011np}). Therefore if the conformal coupling is taken as the
guide to extend the propagation of a scalar field to an arbitrary background,
it would be natural to consider its backreaction in the gravitational field.
Indeed, this scalar scalar field gives place to the
Bocharova-Bronikov-Melnikov-Bekenstein black hole (for references see
\cite{Bekenstein:1996pn}). When the cosmological constant is included in the
gravity sector and a quartic self interaction is included for the scalar field
the Martinez-Troncoso-Zanelli (\textbf{MTZ}) black hole arises
\cite{Martinez:2002ru}. Furthermore, the most general Petrov type D solution
of General Relativity in vacuum, the Plebanski-Demianski family\footnote{With
event horizons and no conical singularities.}, has been shown to be an exact
solution of this system \cite{Anabalon:2009qt} (it reduces in the non-rotating
case to the C-metric like configuration also reported in
\cite{Charmousis:2009cm}). Since anyway the Weyl invariance of the action is
spoiled due to the inclusion of the Einstein-Hilbert term it is interesting to
explore what happens when the scalar field potential is deformed to include
non-Weyl invariant terms. In this paper the inspiration have been taken from a
realistic real scalar field, the $\pi^{0}$, and its linear sigma model
description, the typical soft symmetry breaking linear term have been
included\footnote{The possibility of including this term was pointed out to us
by Alfonso Zerwekh.} as well as a cubic and a quartic selfinteraction
\cite{Donoghue:1992dd}.

The action principle is thus:%

\begin{equation}
S(g,\phi)=\int d^{4}x\sqrt{-g}\left[  \frac{R-2\Lambda}{2\kappa}-\frac{1}%
{2}\left(  \partial\phi\right)  ^{2}-\frac{1}{12}\phi^{2}R-V(\phi)\right]  ,
\label{AP}%
\end{equation}
with field equations:
\begin{equation}
G_{\mu\nu}+\Lambda g_{\mu\nu}=\kappa T_{\mu\nu}, \label{eqs}%
\end{equation}%
\begin{equation}
T_{\mu\nu}=\partial_{\mu}\phi\partial_{\nu}\phi-\frac{1}{2}g_{\mu\nu}\left(
\partial\phi\right)  ^{2}-g_{\mu\nu}V(\phi)+\frac{1}{6}\left(  g_{\mu\nu
}\kern1pt\vbox{\hrule height 0.9pt\hbox{\vrule width 0.9pt\hskip
2.5pt\vbox{\vskip 5.5pt}\hskip 3pt\vrule width 0.3pt}\hrule height
0.3pt}\kern1pt-\nabla_{\mu}\nabla_{\nu}+G_{\mu\nu}\right)  \phi^{2}%
\end{equation}%
\begin{equation}
\kern1pt\vbox{\hrule height 0.9pt\hbox{\vrule width 0.9pt\hskip
2.5pt\vbox{\vskip 5.5pt}\hskip 3pt\vrule width 0.3pt}\hrule height
0.3pt}\kern1pt\phi=\frac{1}{6}R\phi+\frac{\partial V}{\partial\phi},
\end{equation}
where $G_{\mu\nu}$ is the Einstein tensor and $V(\phi)=\alpha_{1}\phi
+\alpha_{3}\phi^{3}+\alpha_{4}\phi^{4}$. Exact, analytical, solutions exist in
the previous system when $\alpha_{4}=-\frac{\kappa\Lambda}{36}$, $\alpha
_{1}=-\frac{6}{\kappa}\alpha_{3}$. Using the convenient parametrization
$\alpha_{3}=-\frac{\Lambda\sqrt{6\kappa}}{9}\frac{\xi}{\xi^{2}+1}$ the
solutions take the following form:%

\begin{equation}
ds^{2}=\frac{\left(  r+\left(  \xi-1\right)  M\right)  ^{2}}{\left(
r-M\right)  ^{2}}\left[  -\left(  k(1-\frac{M}{r})^{2}-\frac{\lambda r^{2}}%
{3}\right)  dt^{2}+\frac{dr^{2}}{\left(  k(1-\frac{M}{r})^{2}-\frac{\lambda
r^{2}}{3}\right)  }+r^{2}d\Sigma_{k}\right]  \label{sol1}%
\end{equation}

\begin{equation}
\phi=\left(  \frac{6}{\kappa}\right)  ^{1/2}\frac{-\xi r+M\left(
\xi-1\right)  }{r+M\left(  \xi-1\right)  },\qquad\lambda=\frac{\Lambda\left(
\xi^{2}-1\right)  ^{2}}{\xi^{2}+1}, \label{sol}%
\end{equation}
where $d\Sigma_{k}$ is the line element of a surface of constant curvature
$k=\pm1$.

The solution (\ref{sol1}) and (\ref{sol}) is real if and only if $\left\vert
\frac{18\alpha_{3}}{\Lambda\sqrt{6\kappa}}\right\vert <1$. Another solution is
generated by the symmetry of the action
\begin{equation}
\phi\rightarrow-\phi,\qquad\xi\rightarrow-\xi. \label{Trans}%
\end{equation}
In what follows the solutions generated by (\ref{Trans}) will be called the
negative branch.

Some generic features of these solutions are (straightforward modifications
extend all these conclusions to the negative branch):

\begin{itemize}
\item There are curvature singularities at $r=0$ and $r=\left(  1-\xi\right)
M$. This last equation also defines the surface where the scalar field is singular.

\item The MTZ \cite{Martinez:2002ru} family of solutions is recovered when
$\xi=0$.

\item If $\xi=-1$ then $\lambda=0$ and the metric is asymptotically flat.
There is a naked singularity at $r=2M$ and at $r=0$. In this case the negative
branch has constant scalar field.

\item The metric (\ref{sol1}) has two asymptotic regions, $r=\infty$ and
$r=M$. The spacetime has a different constant curvature in each of these
boundaries as can be seen from the Riemann tensor $\lim_{r\rightarrow\infty
}R_{\cdot\cdot\lambda\rho}^{\mu\nu}=\frac{\lambda}{3}\delta_{\lambda\rho}%
^{\mu\nu}$, $\lim_{r\rightarrow M}R_{\cdot\cdot\lambda\rho}^{\mu\nu}%
=\frac{\lambda}{3\xi^{2}}\delta_{\lambda\rho}^{\mu\nu}$.

\item It is possible to change $\xi$ by $\xi^{-1}$ with a diffeomorphism. Note
that (\ref{sol1}) and (\ref{sol}) seems to be two solutions, namely each one
of the real roots of the equation
\begin{equation}
\alpha_{3}=-\frac{\Lambda\sqrt{6\kappa}}{9}\frac{\xi}{\xi^{2}+1}, \label{quad}%
\end{equation}
that is $\xi$ and $\xi^{-1}$. However, the map $\xi\rightarrow\xi^{-1}$ on the
configuration (\ref{sol1}), (\ref{sol}) plus the diffeomorphism $r=\rho\xi,$
$t=\xi T$ and the reparametrization $M\rightarrow\xi M$ it is equivalent to
the apply the diffeomorphism
\begin{equation}
r=\frac{\rho M}{\rho-M} \label{C2}%
\end{equation}
to the same configuration. Actually, this diffeomorphism interchange the
location of infinity, the boundary at $r=\infty$ is mapped to $\rho=M$ and the
one at $r=M$ is mapped to $\rho=\infty$.

\item The denominator of the scalar field never vanishes for $\xi>0$ and
$r\geq M$.

\item There is also a region between $0<r<M$ which make sense as a black hole
when the cosmological constant is negative and the horizon is locally
isomorphic to $H^{2}$.

\item The effective potential $W\left(  \phi\right)  =V\left(  \phi\right)
+\frac{1}{12}\phi^{2}R+\kappa^{-1}\Lambda$ of the scalar field is different in
each of the boundaries, the evaluation of its first derivative in the
configuration (\ref{sol1}) and (\ref{sol}) gives:%

\begin{equation}
\frac{dW}{d\phi}=\sqrt{\frac{6}{\kappa}}\frac{2\Lambda Mr\left(  \xi
^{2}-1\right)  ^{3}\left(  r-M\right)  \left(  r-2M\right)  }{3\left(  \xi
^{2}+1\right)  \left(  r+M\left(  \xi-1\right)  \right)  ^{4}}%
\end{equation}
It follows that there is an extremum of the potential at each of the
asymptotic regions. The mass of the scalar field at the critical points is%

\begin{equation}
\lim_{r\rightarrow\infty}\frac{d^{2}W}{d\phi^{2}}=-\frac{2\Lambda\left(
\xi^{2}-1\right)  \left(  2\xi^{2}+1\right)  }{3\left(  \xi^{2}+1\right)
},\qquad\lim_{r\rightarrow M}\frac{d^{2}W}{d\phi^{2}}=\frac{2\Lambda\left(
\xi^{2}-1\right)  \left(  \xi^{2}+2\right)  }{3\xi^{2}\left(  \xi
^{2}+1\right)  },
\end{equation}%
\begin{equation}
\lim_{r\rightarrow2M}\frac{d^{2}W}{d\phi^{2}}=-\frac{4\Lambda\left(
\xi-1\right)  ^{2}}{3\left(  \xi^{2}+1\right)  }\text{.} \label{masses}%
\end{equation}

\item The scalar field acquires a different non trivial vacuum expectation
value at each of the boundaries. For the branch (\ref{sol1}) and (\ref{sol})
$\phi\left(  \infty\right)  =-\left(  \frac{6}{\kappa}\right)  ^{1/2}\xi$,
$\phi(M)=-\left(  \frac{6}{\kappa}\right)  ^{1/2}\frac{1}{\xi}$. Therefore it
is possible to make a field redefinition to have a scalar field that goes to
zero in one of the boundaries but not on both at the same time. It turns out
that the potential written in terms of the field $\psi=\phi-\phi\left(
\infty\right)  $ have no linear term at infinity as can be seen from the fact
that $\psi=0$ is a critical point of it. The same happens at the other
boundary using the field redefinition $\psi=\phi-\phi\left(  M\right)  $.

\item The effective gravitational coupling of this model is dynamical, as can
be seen by direct inspection of the action principle (\ref{AP}). It has
different values at each boundary of the spacetime:
\begin{equation}
\lim_{r\rightarrow\infty}G_{eff}=G\left(  1-\xi^{2}\right)  ,\qquad
\lim_{r\rightarrow M}G_{eff}=G\left(  \xi^{2}-1\right)  \xi^{-2}. \label{C1}%
\end{equation}
However, in a Cavendish experiment the effective coupling is different and it
is given by \cite{Faraoni:2006fx}
\begin{equation}
G_{eff}^{\ast}=2G\frac{1-\frac{\kappa}{9}\phi^{2}}{\left(  1-\frac{\kappa}%
{6}\phi^{2}\right)  \left(  1-\frac{\kappa}{12}\phi^{2}\right)  }.
\end{equation}
At each boundary the following values are taken%
\begin{equation}
\lim_{r\rightarrow\infty}G_{eff}^{\ast}=2G\frac{2\left(  2\xi^{2}-3\right)
}{3\left(  2-\xi^{2}\right)  \left(  \xi^{2}-1\right)  },\qquad\lim
_{r\rightarrow M}G_{eff}^{\ast}=2G\frac{2\xi^{2}\left(  3\xi^{2}-2\right)
}{3\left(  2\xi^{2}-1\right)  \left(  \xi^{2}-1\right)  }.
\end{equation}
It follows that $G_{eff}^{\ast}$ is positive at the boundaries if $\frac{1}%
{2}<\xi^{2}<\frac{2}{3}$.

\item The stress energy tensor of the scalar field is that of an anisotropic
fluid. In an orthonormal frame it has the form of $T^{ab}=diag(\rho
,p_{1},p_{2},p_{2})$. The explicit expression for each of of the components is
rather involved, however some conclusions can be drawn from
\begin{align}
\rho+p_{1}  &  =-\frac{4M\xi(r-M)}{\left(  r+M\left(  \xi-1\right)  \right)
^{4}\kappa}\left(  k(1-\frac{M}{r})^{2}-\frac{\lambda r^{2}}{3}\right)
,\label{A}\\
\rho+p_{2}  &  =-\frac{2Mk(r-M)^{3}\left(  -\xi r+M\left(  \xi-1\right)
\right)  }{\kappa r^{4}\left(  r+M\left(  \xi-1\right)  \right)  ^{3}}.
\label{B}%
\end{align}
From (\ref{A}) and (\ref{B}) it follows that for the black holes neither the
weak nor the null energy conditions can be everywhere satisfied due to the
change of sign of the expression
\begin{equation}
k(1-\frac{M}{r})^{2}-\frac{\lambda r^{2}}{3} \label{exp}%
\end{equation}
in the horizons. For anti de Sitter wormholes (\ref{exp}) it is always
positive and therefore the above mentioned energy conditions do not hold in
the region $r>M>0$ when $\xi>0$. For de Sitter cosmologies (\ref{exp}) it is
always negative thus $\rho+p_{1}\geq0$ and, moreover, $\rho+p_{2}\geq0$ in the
region $r>M>0$ with $\xi>0$. Therefore the null energy condition holds in the
case of the bouncing cosmologies to be discussed in the next section.
\end{itemize}

\section{Geometrical characterization of the solutions.}

\subsection{$\Lambda>0$ and $k=1.$}

In this case it is convenient to set $\lambda=\frac{3}{L^{2}}$. Expression
(\ref{exp}) has three positive roots for $\frac{L}{4}>M>0$:%

\begin{equation}
r_{++}=\frac{L}{2}\left(  1+\sqrt{1-4ML^{-1}}\right)  ,\qquad r_{+}=\frac
{L}{2}\left(  1-\sqrt{1-4ML^{-1}}\right)  ,\qquad r_{-}=\frac{L}{2}\left(
-1+\sqrt{1+4ML^{-1}}\right)  .
\end{equation}
they satisfy $L>r_{++}>\frac{L}{2}>r_{+}>M$ $>r_{-}>0$. It follows that when
$\infty>r>M$ there is a black hole with the conformal structure of the Kottler
solution (also known as the Schwarzschild-de Sitter solution) with the inner
singularity replaced by a new asymptotic region. When $M>r>0$ there is only
the cosmological horizon given by $r_{-}$ and therefore the singularity is
naked. When $M<0$ then $\infty>r>0$ and there is only a cosmological horizon
at $r_{++}$. When $M>\frac{L}{4}$ then $r_{++}$ and $r_{+}$ becomes imaginary
and the region between $\infty>r>M$ becomes an inhomogenous bouncing cosmology
(for a review see \cite{Novello:2008ra}) with the initial condition at $r=M$
being a completely homogenous and isotropic universe with cosmological
constant $\frac{\lambda}{\xi^{2}}$, which after going through an inhomogeneous
phase it ends in a completely homogenous de Sitter universe with cosmological
constant $\lambda$. The bounce is located at
\begin{equation}
r=(1+\sqrt{\xi})M.
\end{equation}

It should be noted that the exact solution presented here allows the Universe
to evolve from a very large cosmological constant (when $\xi\approx0$) to a
very small one. At the same time the scalar field evolves from a local maximum
of the potential to a local minimum. Note also that $\xi$ measures the
deviation from conformality of the matter Lagrangian, so a small deviation
from conformality is required for this phenomenon to exist.

\subsection{$\Lambda>0$ and $k=-1.$}

In this case the metric is everywhere regular, $g_{00}$ never vanishes, and
the spacetime can be interpreted as inhomogenous bouncing cosmologies with the
time given by the coordinate $r\in\lbrack M,\infty]$. The bounce is located at
$r=(1+\sqrt{\xi})M.$

\subsection{$\Lambda<0$ and $k=-1$.}

In this case it is convenient to set $\lambda=\frac{3}{L^{2}}$. Expression
(\ref{exp}) has three positive roots for $\frac{L}{4}>M>0$%

\begin{equation}
r_{++}=\frac{L}{2}\left(  1+\sqrt{1-4ML^{-1}}\right)  ,\qquad r_{+}=\frac
{L}{2}\left(  1-\sqrt{1-4ML^{-1}}\right)  ,\qquad r_{-}=\frac{L}{2}\left(
-1+\sqrt{1+4ML^{-1}}\right)  .
\end{equation}

In this case there is no cosmological horizon and the two asymptotic regions
are separated by two event horizons. All of what it was said in the case of
$\Lambda>0\,\ $and $k=1$ apply in this case changing the cosmological horizon
for a black hole horizon.\thinspace\ When $M>r>0$ there is a black hole with a
single event horizon and singularity at $r=0$. When $M<0\,\ $then $\infty>r>0$
and there is also a black hole with a single event horizon at $r_{++}$. When
$M>\frac{L}{4}$ there is a wormhole with the throat located at $r=(1+\sqrt
{\xi})M.$ The wormhole interpolates between two asymptotically locally anti de
Sitter regions with a different value for the cosmological constant ($\lambda$
and $\frac{\lambda}{\xi^{2}}$). The scalar field is everywhere regular
whenever $M>0$ and $\xi>0$.

\subsection{$\Lambda<0$ and $k=1.$}

This case represent wormhole solutions with the boundary in the conformal
class of either $R\times S^{2}$. The throat is located at $r=(1+\sqrt{\xi})M$
and the metric interpolates between two asymptotically anti de Sitter regions
with a different value for the cosmological constant ($\lambda$ and
$\frac{\lambda}{\xi^{2}}$). The configuration is everywhere regular whenever
$M>0$ and $\xi>0$.

\section{Final Remarks.}

A new class of exact solutions has been presented in this paper. Some of them
have the interesting property of having no singularity at all, neither in the
spacetime manifold nor in the matter configuration. The singularity inside the
black hole is replaced by another asymptotic region, implying, when the
horizons are removed, the existence of a new set of either wormholes or
bouncing cosmologies. These are the first, exact, asymptotically anti de
Sitter wormholes, bouncing de Sitter cosmologies and regular black holes in
four dimensions for the Einstein-conformally coupled scalar field system.

The model discussed here (\ref{AP}) looks rather particular. In principle, it
would be desirable to have a classification of all the possible potentials
that are compatible with the Einstein equations. Indeed, since the Einstein
equations are non-linear one could expect to find that within certain Petrov
class of metrics some restrictions on the scalar potential should hold. This
is actually the case as will be reported in a forthcoming article.

Finally, to name some open questions let by this work we would like to mention
its extension to include the Maxwell field. Another follow up is to study
either thermodynamical properties or stability of the solutions as was done
for the MTZ black hole in \cite{Barlow:2005yd} and \cite{Harper:2003wt} respectively.

%%%%%%%%%%%%%%%%%%%%%%%%%%%%%%%%%%%%%%%%%%%%%%%%%%%%%

\end{document}